\documentclass{iau}
\usepackage{graphicx}
\usepackage{natbib}

\title[PlanetPack] 
{PlanetPack software tool for exoplanets detection: coming new features}

\author[Roman V. Baluev]   
{Roman V. Baluev$^{1,2}$}

\affiliation{$^1$Central Astronomical Observatory at Pulkovo of Russian Academy of Sciences,
Pulkovskoje shosse 65, St Petersburg 196140, Russia \\[\affilskip]
$^2$Sobolev Astronomical Institute, St Petersburg State University, Universitetskij
prospekt 28, Petrodvorets, St Petersburg 198504, Russia \\ email: {\tt r.baluev@spbu.ru}}

\pubyear{2014}
\volume{310}  
\pagerange{xxx--xxx}
\jname{Complex Planetary Systems}
\editors{Z. Knezevic \& A. Lema\^itre}
\begin{document}

\maketitle

\begin{abstract}
We briefly overview the new features of PlanetPack2, the forthcoming update of PlanetPack,
which is a software tool for exoplanets detection and characterization from Doppler radial
velocity data. Among other things, this major update brings parallelized
computing, new advanced models of the Doppler noise, handling of the so-called Keplerian
periodogram, and routines for transits fitting and transit timing variation analysis.
\end{abstract}

\firstsection 
\section{Introduction}
PlanetPack is a software tool that facilitates the detection and characterization of
exoplanets from the radial velocity (RV) data, as well as basic tasks of long-term
dynamical simulations in exoplanetary systems. The detailed description of the numeric
algorithms implemented in PlanetPack is given in the paper \citep{Baluev13}, coming with
its initial 1.0 release. After that several updates of the package were released, offering
a lot of bug fixes, minor improvements, as well as moderate expansions
of the functionality. As of this writing, the current downloadable version of PlanetPack is
1.8.1. The current source code, as well as the technical manual, can be downloaded at
\texttt{http://sourceforge.net/projects/planetpack}.

Here we pre-announce the first major update of the package, PlanetPack 2.0, which should be
released in the near future. In addition to numerous bug fixes, this update includes a
reorganization of the large parts of its architecture, and several new major algorithms.
Now we briefly describe the main changes.

\section{PlanetPack2: transits fitting and other new features}
The following new features of the PlanetPack 2.0 release deserve noticing:
\begin{enumerate}
\item Multithreading and parallelized computing, increasing the performance of
some computationally heavy algorithms. This was achieved by migrating to the new
ANSI standard of the C++ language, C++11.

\item Several new models of the Doppler noise can be selected by the user, including
e.g. the regularized model from \citep{Baluev14}. This regularized model often helps
to suppress the non-linearity of the RV curve fit.

\item The optimized computation algorithm of the so-called Keplerian periodogram
\citep{Cumming04}, equipped with an efficient analytic method of calculating its
significance levels (Baluev 2014, in prep.).

\item Fitting exoplanetary transit lightcurves is now implemented in PlanetPack.
This algorithm can fit just a single transit lightcurve, as well as a series of transits
for the same star to generate the transit timing variation (TTV) data. These TTV
data can be further analysed as well in order to e.g. reveal possible periodic
variations indicating the presence of additional (non-transiting) planets in the system.
The transit lightcurve model is based on the stellar limb darkening model by
\citep{AbuGost13}. Also, the transit fitting can be performed taking
into account the red (correlated) noise in the photometry data.
\end{enumerate}

\section{Plans for future work}
Some results of the PlanetPack TTV analysis of the photometric data from the Exoplanet
Transit Database, \texttt{http://var2.astro.cz/ETD/}, will be soon presented in a separate
work. Concerning the evolution of the PlanetPack code, we plan
to further develop the transit and TTV analysis module and to better integrate it with the
Doppler analysis block. We expect that in a rather near future PlanetPack should be able to
solve such complicated tasks as the simultaneous fitting of the RV, transit, and TTV
data for the same star. This integration should also take into account subtle intervenue
between the Doppler and photometry measurements like the Rositter-McLaughlin effect.

\medskip
{\small The work was supported by the President of Russia grant for young scientists
(MK-733.2014.2), by the Russian Foundation for Basic Research (project 14-02-92615 KO\_a),
and by the programme of the Presidium of Russian Academy of Sciences
``Non-stationary phenomena in the objects of the Universe''.}

\end{document}